James Brusseau, Philosophy, Pace University NYC
jbrusseau@pace.edu

# What to Do When Privacy is Gone

## Abstract

Today's ethics of privacy is largely dedicated to defending personal information from big data technologies. This essay goes in the other direction. It considers the struggle to be lost, and explores two strategies for living after privacy is gone. First, total exposure embraces privacy's decline, and then contributes to the process with transparency. All personal information is shared without reservation. The resulting ethics is explored through a big data version of Robert Nozick's Experience Machine thought experiment. Second, transient existence responds to privacy's loss by ceaselessly generating new personal identities, which translates into constantly producing temporarily unviolated private information. The ethics is explored through Gilles Deleuze's metaphysics of difference applied in linguistic terms to the formation of the self. Comparing the exposure and transience alternatives leads to the conclusion that today's big data reality splits the traditional ethical link between authenticity and freedom. Exposure provides authenticity, but negates human freedom. Transience provides freedom, but disdains authenticity.

**Privacy Gone**

Privacy invasions by social media platforms and consumer marketing enterprises encounter increasingly militant resistance. But what if it's too late? What if resistance won't work because we've already lost the struggle to control our personal information? (Brusseau 2019) As early as 1999 Scott McNealy, head of Sun Microsystems, suspected resistance would prove futile when he counseled, "You have zero privacy anyway. Get over it." (Sprenger 1999) Either he was prescient or wrong, and if he wasn't wrong then the ethics of big data swings away from defending privacy, and toward living without it.

Here are two ways of living without it. One embraces privacy's loss, the other creates a new privacy to replace the one we used to have.

**Exposure**

Privacy's absence can be embraced as exposure. Who we are at every time of the day and night is displayed, unreservedly and indiscriminately.

As a theoretical condition, personal transparency articulates easily, but in practice it's more difficult to envision. Cameras and microphones everywhere is a start, including those focused on reading emotions from muscle twitches in the face, but that's only an initiation. Memories we may want to forget, aspirations we would prefer to disavow, urges we may not want to acknowledge, all those too become publicly available. As they do, the idea of exposure slips from interesting into revolting.

But it's also enticing. If revealing even our most fleeting and remote desires to the matchmakers at Tinder will open passion on demand, then more than a few are going to be tempted. And, while no one wants to answer the question *Why do you want this job?* with unvarnished truth, if the information is transmitted to the recruiter LinkedIn, which is empowered by data and algorithms to industriously deliver tantalizing professional opportunities from around the globe, then maybe the gain compensates the exposure.

As technologies including human tagging advance (Voas & Kshetri 2017), transparency's rewards surge. How much will busy parents pay – in terms of intimate personal information – to find a vacation that brings their family together? How much unfiltered biological data will patients transmit to healthcare enterprises – perhaps via an implanted microchip – for a guaranteed alert one hour before a heart attack? (Raghupathi & Raghupathi 2014)

If the answer is everything, the codes and charts of our bodies, the full array of our memories, the whole psychological truth of our desires, then we're exposed.

The first effect of exposure is that we're perfectly integrated as selves in this sense: no aspect of our identity can be compartmentalized for selective distribution. I mean, one way I stay out of jail is by presenting – and equally by concealing – aspects of myself. A father in the morning, a lecturer in the afternoon, a husband in the evening. If I treated my children as colleagues, I'd fail as a parent. If I approached the women at work as my wife, I'd end up incarcerated. So, identity today is about visibility *and* concealment. That distinction collapses, though, in the coming reality of transparency because everyone sees all the way through everyone else.

Several years ago news reports circulated about a woman who lived nocturnally as a sex worker, while maintaining an ordinary daytime identity with an academic email address and typical social media postings. The two worlds kept their distance, until she and her clients began appearing in each other's "People You May Know" recommendations on Facebook. (Hill 2017) She tried to turn off the connections with the expected results, and so learned first-hand two critical aspects of contemporary reality: a normal day features nearly schizophrenically diverse personalities, and normal schizophrenia is jeopardized – for the first time in history – by the big data invasion of privacy.

The second transparency effect is fulfilled authenticity. As delineated by the tradition from Kierkegaard to Heidegger, the struggle to be ourselves involves divining our own projects and endeavors amidst the distractions of quotidian life. Advertisers leverage psychological expertise to shape our wants, celebrities exploit fame to show us how we should live, coworkers engulf our aspirations with theirs. The authenticity challenge has always been to screen out these diversions and get through to our own being in the world. The challenge is no longer even a minor difficulty, however, since everything there is to know for each of us about who we are, on all levels, is revealed to everyone, including ourselves. Of course there's no guarantee that the exposed will actually *live* the projects that are revealed as theirs, but they will be unable to avoid facing them.

There's a connection between big data ethical reality and Robert Nozick's Experience Machine thought experiment from the 1970s, about a decade before Apple computers went mainstream. What Nozick imagined was a was a floating tank of warm water, a sensory deprivation chamber where electrodes wrapped around our heads to feed our synapses a tantalizing experience indistinguishable from lived reality. Would you trade, the thought experiment asked, your outside life for existence inside the tank, one that would be as thrilling or heroic or luxurious as you wish, but only in your mind? It's a hard call. What's at stake, though, is easy to see. It's mental sort of hedonistic happiness versus a tangible personal freedom: you get prefabricated episodes guaranteed to feel good, while giving up the possibility of *creating* experiences and an identity for yourself out in the unpredictable world.

Coming post-privacy reality offers an analogous choice in this sense: going inside promises satisfactions, but implies relinquishing control over our own destinies. When we are radically exposed, and so entirely known by big data economic ventures, the provided services reach a perfection that converts into confinement. As a crude example, Netflix intersects its users' personal information with predictive analytics to begin rolling the next film before the previous ends: you get what you want *before* making any choices at all. On one level, you don't select another movie from a list and, above that, you don't even decide whether to watch another because it's *already* going.

One series that has been selected for me is Black Mirror, which includes an episode depicting a couple in a restaurant getting served their dishes just as they are about to ask to see the menu. In a big data future of embraced transparency, the hyper efficiency shouldn't be disconcerting. And it won't only be movie selections and dinner choices. Every need and want will be answered so immediately that we won't have time to understand *why* it's right, *when* we started wanting it, or even to ask what it is that we wanted in the first place.

We're not choosing anymore, or even choosing to not choose.

A curiosity about life in the Big Data Experience Machine is that the way we realize something is what we want is: we already have it. And that's the *only* way we know we want something. More, if we *do* feel the urge for a slice of pizza, or to binge Seinfeld, or to incite a romantic fling, what that really means is: we *don't* actually want it. We can't since the whole idea of the functioning mechanism is that it knows us transparently and so responds to our desires so perfectly that they're answered without even a moment of suffering an unfulfilled craving.

The third transparency effect, consequently, after personal integrity and manifest authenticity, is unimaginable convenience, *literally* unimaginable. It is a delivered and tranquilizing enjoyment so perfectly delineating our integrated, authentic selves that it must be experienced to be understood. More, thinking about it ruins the experience because it introduces *doubt*, the possibility that what we're receiving is not quite right, or arrived just a moment late. The power of the big data version of the experience machine, however, is precisely the irrecoverable deletion of those concerns. So, it's not just that the joys must be experienced to be understood, they can *only* be experienced to be understood.

If we had more time we could ask whether something truly satisfies if we get it before realizing a hunger for it, but no matter the answer, the drowning of personal desire in convenience and delight is not a criticism of the big data experience machine, it's a temptation. And also one that implies a counter-intuitive exchange. We get personal authenticity, a life of satisfactions that perfectly incarnates who we are. It's also true though, that we have no personal freedom to determine who that someone is. Since there's no experimenting with new possibilities, or struggling with what's worth pursuing and having, there's no room for creating an identity. There's only room to receive the bliss that initially recommended we fully expose ourselves to big data reality. Personal transparency means a life so ecstatically ours – so authentic – that we can't do anything with it.

Viewed from outside, exposure and the end of privacy is a digital Stockholm syndrome. By revealing their personal information, those who have chosen transparency embrace the information sets and algorithms that control their experiences: those whose freedom has been arrested are actually *grateful* to their captors. But, from *within* the experience, the very idea of being captive doesn't make sense since it's impossible to encounter – or even conceptualize – any kind of restraint. If we always get everything we want – movies, dinners, jobs, lovers, everything – before we even know that we want them, how could we feel anything but total liberation?

**Transience**

The second response to privacy's loss is to cancel it by changing who we are.

The strategy is to develop new traits to replace the personal information that has been gathered to label and contain our identity. When that happens, privacy's loss stops mattering because we are no longer the people whose data has been exposed. Stronger, it's not just that the loss stops mattering, it's that the idea itself of privacy's violation no longer makes sense since there's no one left to suffer the transgression.

Identity transformations sufficiently powerful to crack big data profiles are common, if not frequent. People recreate who they are seismically when they go away to college, marry, have children. Think of the kinds of things people say, do, and want around 3 a.m. at each of those stages. The differences measure how badly the connections and incentives aimed by social media platforms and marketers suddenly begin to miss as individuals cross the thresholds. It's not just that they're off-target, it's that they aim for people who are no longer there.

There are also more surreptitious opportunities for immunity from our own pasts, some hiding underneath the very platforms and technologies that threaten to build inescapable identity data banks. Take LinkedIn. Designed to harvest personal information and then display job openings curated for information-modelled identities (Staddon 2009), it doesn't just offer, it also excludes, it methodically eliminates opportunities in the name of advancing careers along established lines. So, nurses receive leads for new posts at other hospitals and in elder care businesses and, just as industriously, get shielded from new openings in accounting departments and at law firms. With a little search ingenuity, though, users can discover options their own past information would otherwise prohibit. The Silicon Valley entrepreneur may locate a corporate post in the Swedish welfare state, the feminist can purchase a headscarf and a one-way ticket to an English-teaching job in Saudi Arabia. Regardless, life-jarring opportunities are out there. Personal information can be hacked to cross the wires of LinkedIn's algorithms, and so produce deviant possibilities, and for those who transmit a resonant appeal to the recruiter who's willing to take a chance, they can be gone the next day.

Platforms designed to generate music playlists can be twisted, romantic matchmaking applications can be perverted. The list of possibilities for generating encounters entirely detached from familiar habits and attractions go on as long as the app store catalog. So, while it's true that it has never been harder to *not* be who we are because the identity we've compiled digitally keeps tagging along behind us, it has also never been easier to *get out* of who we are, to disrupt our existences from the bottom up by connecting with and nurturing unfamiliar tastes, urges, and potentials. When that happens, lives regenerate with re-established zones of personal data, ones that haven't been scraped and commodified for the information marketplace.

At least not yet. The scraping and commodification will come again, though, because big data reality means the information merchants will always be catching up. People can change who they are, they can recreate their defining traits to establish an unexpected zone of privacy, but the data gatherers can also keep searching phone records and tracking credit card histories and scanning social media accounts to continue resolving, inexorably, their personal information profiles.

To stay ahead of the surveillance and the big data capitalists, what's needed is not just identity generation, but ceaseless regeneration. Consequently, if the first transience effect is identity production, the second is repetition in the excessive, Bataillian sense (Bataille 1991). The work of becoming other must repeat, and perpetuate itself.

*In*authenticity, that means, becomes a guiding virtue. Since the endeavor is to escape our own personal information by reassembling it elsewhere and differently and constantly, the only reason for consolidating a stable and genuine sense of who we are at any one time and place is *so that* we may become someone else. Identity is no longer a termination but a station, instead of an end in itself, it serves comings and goings. Where the Heideggerian vision organizes human experience around the project of discovering a durable self-understanding in spite of a jagged and distracting world, the idea here is to wield those deviant possibilities to cut away from established self-understandings. For example, a job announcement discovered by chance and pursued on a whim may be irreconcilable with posts we have previously held, and may demand relocation to a foreign place. The response to those uncertainties is not contemplative, but abrupt: buy a one-way airline ticket. Inauthenticity is the enabling attitude, and because identity is conceived to maximize transience, there's an accompanying redefining of superficiality and impulsiveness.

When it comes to careers, interests, values, and decisions, they are no longer character defects, but powers.[1]

A theory of disrupted selves intersects with Gilles Deleuze's metaphysics of difference when applied to the linguistics of personal identity. The critical move reverses the conventional privileging of nouns over verbs. Nouns as primary means there's someone who I am, and that determines what I do. So, I once lived in Mexico, which means I *am* the sort of person who would go to live there, and that explains why I went. The same episode can be read the other way, however: I went to live there, and consequently *became* that sort of person. (Deleuze 1983) If that's the order – if it's the verb of what is done before the noun of who I am – then what it means to be me or you is an after-effect of incarnating that person. Identity is the result of action, not the other way, which means that we can disassociate from our established selves by engaging in unfamiliar experiences. Going out into the world and doing something different – something irreconcilable with who you are – generates a different identity.

It happens every summer that people depart for backpacking or bicycling trips abroad and intuitively discover that they can create themselves as whomever they wish for encountered strangers. A name, a home, a career aspiration, or romantic inclination, all these things can be invented on the spot and without social penalty. The fact that no one is running background checks on their fellow night-train riders doesn't automatically convert everyone into vivid explorers of experiences they wouldn't engage were their friends watching, but every season there are a few who begin with the identity experiments and destabilizations, and then cut away entirely. Maybe they meet someone who engages with an unfamiliar language, or find a collective organized around an alien value hierarchy. Whatever the particulars, the person who emerges connects only tenuously with the one who was there before. Exotic examples include Paul Bowles's *Sheltering Sky*, or the case of Isabelle Eberhardt, but there's no need to get so extreme. If you visit the travel section of the local bookstore you'll find volume after volume written by people you've never heard of, all telling the story about going abroad and converting into someone whose new personal data can't be resolved with the old (Colley 2000).

From this deeply inauthentic kind of existence, a narrow and irregular human freedom emerges. More exploratory than selective, personal autonomy in the time of big data is not about choosing between known options so much as deciding between the collection of known options, and the unknown. When the libertarian entrepreneur goes corporate, or the feminist goes to Arabia, these aren't weighed possibilities selected from a set of predictable futures. They are blind leaps toward the unforeseeable, and inspired by the valorization of inauthenticity. And since the entire point is to recreate oneself in unexpected ways that escape all the personal information that has been gathered to capture who we are, it's an odd sort of personal freedom by definition because it aims to be impossible – and to make it impossible – to locate the person who experiences it.

**Postscript on big data, freedom, and authenticity**
Ethics from Kierkegaard to Heidegger conceived freedom *through* authenticity. Freedom means autonomy, self-rule, and the purpose of the rule is to discover and to express an enduring self beneath the distractions of modern life. The freedom-authenticity link breaks under pressure of big data reality, though. When privacy is gone, and when all that remains is the choice between exposure and transience, then we are caught in an either/or divide: either authenticity without freedom, or freedom without authenticity.

**End**

**Notes**
1. There's a test of machine learning or artificial intelligence here. While predictive analytics may be able to foresee what we want today, could the leap be made to the next level to predict what we will *convert* into wanting? In other words, could the machine be adequate to not just who we are, but to how we could recreate ourselves?